# Cultural Preferences to Color Quality of Illumination of Different Objects

Anqing Liu,[1,*] Arūnas Tuzikas,[2] Artūras Žukauskas,[2] Rimantas Vaicekauskas,[3] Prančiskas Vitta,[2,3] and Michael Shur[1,4]

[1] Physics Department, Rensselaer Polytechnic Institute, 110 8th Street, Troy, New York 12180, USA
[2] Institute of Applied Research, Vilnius University, Saulėtekio al. 9-III, LT-10222 Vilnius, Lithuania
[3] Department of Computer Science, Vilnius University, Didlaukio g. 47, LT-08303 Vilnius, Lithuania
[4] Department of Electrical, Computer and System Engineering, Rensselaer Polytechnic Institute, 110 8th Street, Troy, New York 12180, USA
*liua4@rpi.edu

**Abstract:** The preferences to color quality of illumination were investigated for American and Chinese subjects using a solid-state source of white light with the continuously tunable color saturation ability and correlated color temperature of quadrichromatic blends. Subjects were asked to identify both "most natural" and preferred blends. For very familiar objects, cultural differences did not affect the average of the selected blends. For less familiar objects (various paintings), cultural differences in the average selected blends depended on the level of the familiarity of the content. An unfamiliar painting also showed preferences to color temperature being dependent on the cultural background. In all cases, the American subjects exhibited noticeably wider distributions.

## 1. Introduction

Cross-culture studies indicate that different cultural groups show different responses and preferences to lighting conditions [1], colors [2], and color combinations [3,4]. However, little is known about cultural differences in assessing color rendering quality of illumination. Until recently, such needs were difficult to examine because of the lack of light sources with controllable color rendition properties.

The emergence and commercialization of illumination-grade light-emitting diodes (LEDs) has led to the development of solid-state sources of light with the spectral power distributions (SPD) that meet various needs in the color quality of lighting [5,6]. In particular, such SPDs can have very different ability to saturate and desaturate the colors of illuminated objects [7-10]. Moreover, intelligently controlled polychromatic LED clusters can generate metamers of white light with continuously tunable color rendition properties [11]. Such light sources can find many important applications, including improving visual impression of artistic paintings [12-14]. They can meet specific color-quality requirements of people with different subjective needs, including those depending on cultural background.

The subjective perception of colorfulness is also influenced by the content of the illuminated scene and is known to benefit from the image saturation level, especially for familiar objects, since the familiarity contributes to color adaptation [15]. Hence, the subjective need in color rendering quality of illumination depends on the character of the illuminated object and its familiarity to the observer.

In this work, we use a solid-state source of light with tunable color-rendition properties for the investigation of the preferences to the color quality of illumination for subjects from different cultural groups.

## 2. Experimental

The light source used in this work is a color rendition engine described in Ref. 11. The engine comprises red (624 nm peak wavelength), green (523 nm), and blue (450 nm) direct-emission LEDs and phosphor-conversion amber (591 nm) LEDs. The color rendition properties of the engine are tuned by composing the SPD as a weighted sum of the RGB (red-green-blue) and AGB (amber-green-blue) blends, which have high ability to saturate and dull colors, respectively:

$$S_{RAGB}(\lambda) = \sigma S_{AGB}(\lambda) + (1-\sigma)S_{RGB}(\lambda). \quad (1)$$

Provided that the two trichromatic lights have the same correlated color temperature (CCT) and luminous output, tuning the weight parameter $\sigma$ within the interval of (0,1) allows for continuously traversing all possible metameric tetrachromatic (RAGB) blends, including those with the highest color fidelity. The endpoints, $\sigma = 0$ and $\sigma = 1$, correspond to the most color saturating (RGB) and most dulling (AGB) light, respectively.

The color rendition properties of light generated by the engine were quantified using the statistical metric [16], which is a more advanced approach in comparison with the outdated common Color Rendering Index (CRI) measure of the quality of light [17]. The statistical metric is based on sorting 1269 Munsell test color samples by the color shifts within a color space scaled by MacAdam ellipses [18], when the light source under test is replaced by a reference light source. This approach allows for the introduction of different color rendition indices, which are the percentage of colors that are rendered either with high fidelity or with increased or reduced saturation. The most important statistical indices are the Color Fidelity Index (CFI), Color Saturation Index (CSI), and Color Dulling Index (CDI).

Figure 1(a) shows the SPD of the engine adjusted to the highest color fidelity at a CCT of 3000 K. The high-fidelity blend is attained at $\sigma = 0.67$ and has CFI of 87% with low CSI and CDI. Figure 1 (b) shows the corresponding distributions of the color-shift vectors in respect of a blackbody radiator estimated for 218 Munsell samples of value /6, respectively [16]. These distributions are presented within the $a^*$–$b^*$ chromaticity plane of the CIELAB color space; the arrows schematically show the color-shift vectors that are estimated to have perceptual noticeable chromaticity distortions and the open points show chromaticities of colors that are rendered with high fidelity (i.e. their color-shift vectors reside within 3-step MacAdam ellipses and the lightness is altered by less than 2%). When the AGB vs. RGB weight is reduced, the percentage of colors rendered with increased saturation increases and the number of colors rendered with high fidelity decreases. Figures 1(c) and (d) show the SPD and the distribution of the color-shift vectors, respectively, for the engine tuned to $\sigma = 0.36$. At this value of $\sigma$, the CFI drops to 18% and the CSI increases to 68% with the CDI still being low. The most color saturating blend ($\sigma = 0$) has a CSI of 80% and the most color dulling blend ($\sigma = 1$) has a CDI of 64%.

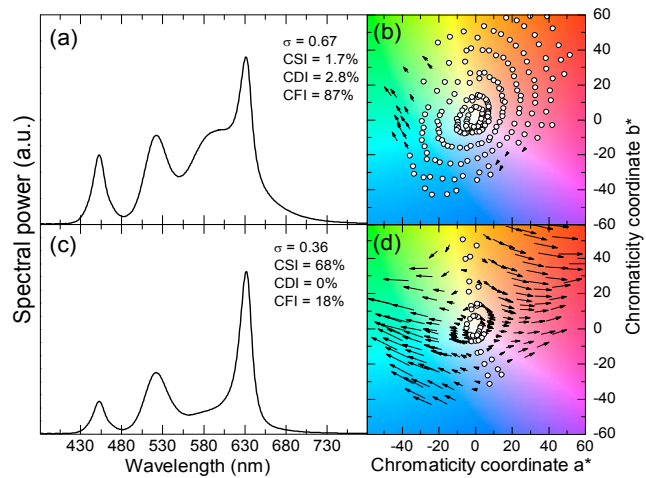

FIG. 1. Spectral power distributions (a) and (c) and distributions of the color-shift vectors (b) and (d) of the color rendition engine adjusted for different dulling vs. saturation weight parameters at a CCT of 3000K. (a) and (b) high-fidelity blend; (c) and (d) a blend with an increased number of colors rendered with increased saturation and reduced number of colors rendered with high fidelity. In Figs. (b) and (d), the arrows schematically show the vectors that are estimated to have perceptually noticeable color distortions and the open points show chromaticities of colors that are rendered with high fidelity.

The experiment was conducted in a dark room without daylight illumination. The engine was mounted on a plastic cabinet pointing downward. The luminance was fixed at 300 lx for all SPDs. The cabinet was sprayed with matt white paint inside to obtain good mixing of light in the interior space. The opening of the cabinet was 50×50 cm in size. Subjects sat ~30 cm away from the cabinet to watch the scene. A computer with a bluetooth interface controlled the engine. The repetitive pressing 'Ctrl'+'→' or

'Ctrl'+'←' resulted in the cyclical incrementing of the weight parameter $\sigma$ from 0 to 1 and then from 1 to 0, respectively, with an increment of 0.05 (gradually traversing 21 different blends ). The correlated color temperature (CCT) was changed from 2500 K to 7500 K and back from 7500 K to 2500 K with an increment of 500 K by pressing 'Ctrl'+'↑' or 'Ctrl'+'↓', respectively. The subjects were able to tune AGB vs. RGB weight and CCT back and forth until a particular task was attained. Then the values of the parameters were recorded by the experimenter (the monitor of the computer was dimmed and kept out of subjects' sight).

205 subjects with normal color vision were selected by Stilling's pseudoisochromatic plate tests and invited to take part in the experiment (see Table 1). 101 individuals were originated from People's Republic of China (they studied and lived in the US for less than a few years) and 104 individuals were locals of the United States. Of total number of subjects, 49 were females and 156 were males with the age ranging from 18 to 33 (the average age of 22). Most of the observers were students or staff and faculty members at Rensselaer Polytechnic Institute.

None of subjects initially possessed detailed knowledge or understanding of the research. Before the experiments, the operating engine was demonstrated to subjects by illuminating an orange by light with different values of the weight parameter $\sigma$ and different CCTs. The observers were introduced to the concepts of natural, saturated, and dull colors as well as of color temperature. The introduction was given in Chinese language for the Chinese observers and in English for the US subjects.

**Table 1. Characteristics and number of observers.**

| Cultural background | Males | Females | Sum | Age range (average) |
|---|---|---|---|---|
| American | 79 | 25 | 104 | 18-32 (23) |
| Chinese | 77 | 24 | 101 | 19-33 (22) |
| Total | 156 | 49 | 205 | 18-33 (22) |

Two types of the experiments were carried out. In the first experiment, we illuminated very familiar objects. Natural fruits and vegetables with different dominant colors were used, including a red apple, a tomato, an orange, two bananas, a lemon, a greenish cucumber and an onion with purple peel. The fruits and vegetables were obtained from a local market. They were being replaced by those with similar appearance every three days to avoid colors variation due to aging.

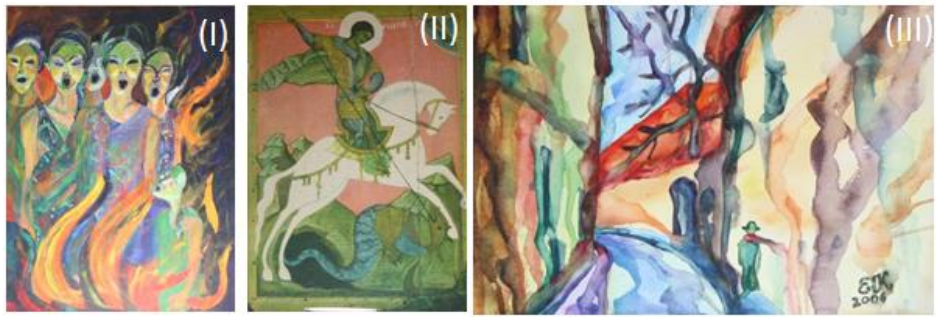

FIG. 2. Three examined paintings that are assumed to have different level of familiarity to investigated subjects: Fire theme (I), Saint George fighting a dragon (II), and Vilnius downtown (III).

In the second experiment, we illuminated unfamiliar objects, which were three paintings (Fig. 2). The first painting (Fig. 2(I); unknown author) was painted by acryl on canvas. It depicted a colorful fire theme that was assumed to be equally unfamiliar to subjects with different cultural backgrounds. The second artwork (Fig. 2(II); unknown author) was painted on a cardboard. It depicted Saint George fighting a dragon in somewhat faded colors. This topic was assumed to be more familiar to subjects with

Christian or multicultural background such as Americans rather than to Chinese. The third artwork used in the experiment (Fig. 2(III); by E. Kuokštis) was a post-impressionistic watercolor painting displaying a medieval downtown scene in Vilnius, Lithuania. The theme of the painting was assumed to be unfamiliar to all subjects, since none of them have been to Vilnius nor were acquainted with the style of the artist.

## 3. Results

In the first experiment (on very familiar objects, i.e. fruits and vegetables), subjects were asked to establish SPDs that rendered colors i) in the "most natural" way (with high fidelity) and ii) in the most appealing way (with the highest subjective preference). For specificity, only the AGB vs. RGB weight was allowed to tune in this experiment, while the CCT was maintained at a value of 3000 K (such CCT corresponds to halogen incandescent lamps that are common in residential lighting). The solid and dashed lines in Fig. 3(a) show the variation of the statistical color quality indices and general CRI of the engine with AGB vs. RGB weight at CCT of 3000 K and 6500 K, respectively.

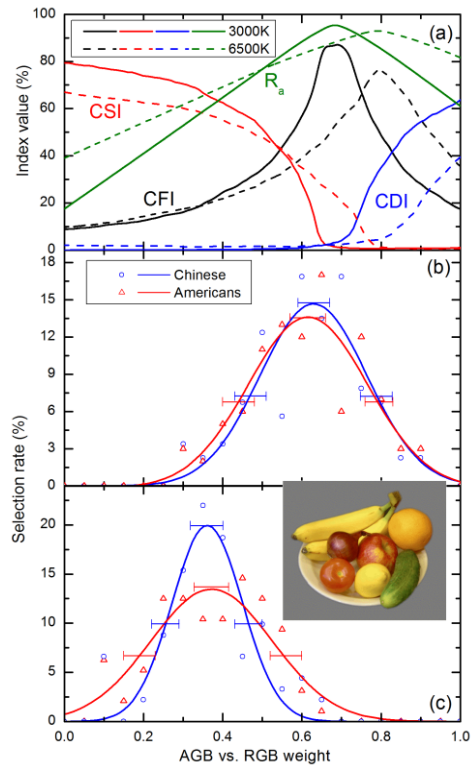

FIG. 3. (a) Statistical color rendering indices and general CRI of the engine as functions of AGB vs. RGB weight for CCTs of 3000 K (solid lines) and 6500 K (dashed lines). (b) and (c) Points, percentage of the "most natural" and preferred subjective selections of the weight parameter $\sigma$, respectively, for the illumination of very familiar objects. Circles and red triangles, data obtained from Chinese and American subjects, respectively. Lines, the Gaussian distributions obtained by least-squares fitting. Horizontal bars, the 95% confidence intervals for the peak weights and FWHMs of the distributions.

Figure 3(b) displays the selection rate for high fidelity blends as a function of the weight parameter $\sigma$. The blue circles and red triangles show the experimental results for Chinese and American subjects, respectively. The solid lines are the least-squares fit of the experimental results to Gaussian distributions. The horizontal bars show the 95% confidence intervals for the peak position and width of the distributions derived from the fitting procedure. The distributions peak at $\sigma_p = 0.63 \pm 0.04$ and $\sigma_p = 0.64 \pm 0.04$ for the American and Chinese groups of subjects, respectively, i.e. no cultural differences in finding the "most natural" conditions of illumination for familiar objects, such as fruits and vegetables, were observed. Moreover, the FWHMs of the measured selection rate distributions are identical for the both groups of subjects (0.35±0.07 and 0.33±0.07, respectively). It is to be noted that the peak selection rates for the both groups of subjects match the highest value of the calculated CFI at $\sigma = 0.67$ within the experimental uncertainty and also agree with the highest general CRI ($\sigma = 0.67$).

Figure 3(c) displays the corresponding results for the selection rate for preferential blends. The peak positions of the distributions of the blends that render the colors of familiar objects with subjective preference are seen to be independent of subjects' cultural background ($\sigma_p = 0.37 \pm 0.04$ and $\sigma_p = 0.36 \pm 0.04$ for the Americans and Chinese, respectively). These distributions are clearly shifted to lower values of the AGB vs. RGB weight that correspond to the blends that render colors with somewhat increased saturation rather than to high-fidelity blends. An important feature of the selection of preferential blends is that the FWHM of the distributions is significantly different for the two groups of subjects. It amounts 0.37±0.07 and 0.21±0.07 for the American and Chinese groups of subjects, respectively.

In the second experiment (on paintings with different level of familiarity), subjects were asked to tune both the AGB vs. RGB weight and CCT. The only task was to establish SPDs of light that renders colors in the most appealing way (i.e. with the highest subjective preference) for each of the three paintings. It is to be noted that at different CCTs, the statistical color quality indices of the engine depend on AGB vs. RGB weight in very similar way, although the dependences are somewhat shifted in respect of each other (cf. statistical color quality indices as functions of AGB vs. RGB weight at different CCTs in Fig. 3(a)).

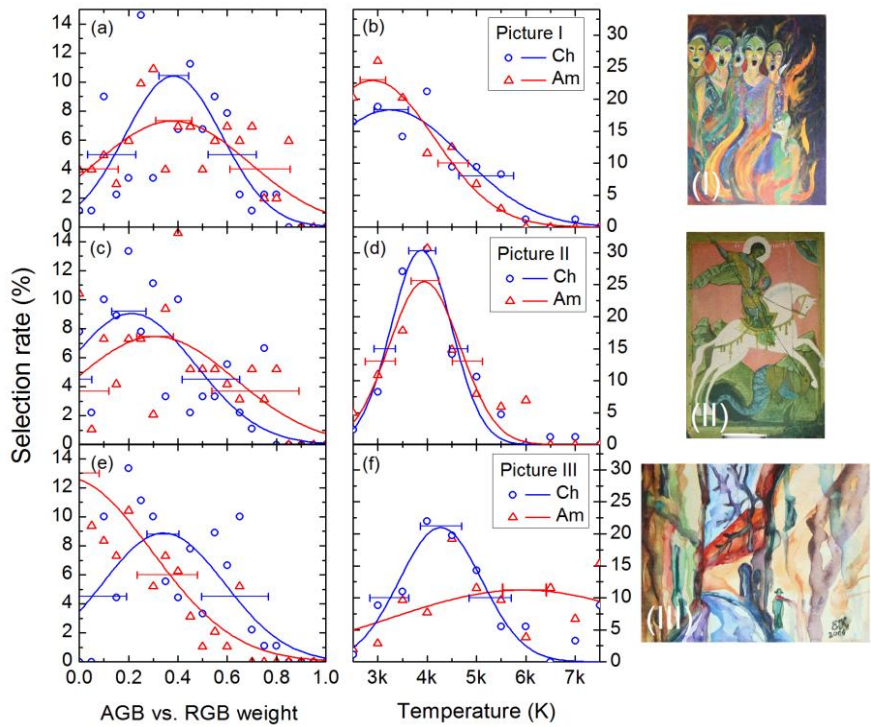

FIG. 4. Points, percentage of the subjective selections of the weight parameter $\sigma$ (a, c, and e) and CCT (b, d, and f) for the most preferred illumination of paintings with various level of familiarity. Blue circles and red triangles, data obtained from Chinese and American subjects, respectively. Lines, Gaussian distributions obtained by least-squares fitting. Horizontal bars, the 95% confidence intervals for the peak weights and FWHMs of the distributions.

Figure 4 shows the experimental results on finding the preferential lighting for the paintings shown in Fig. 2. Figures 4(a), (c), and (e) and Figs. 4(b), (d), and (f) demonstrate the selection rates as functions of the AGB vs. RGB weight and CCT, respectively. The points and lines in Fig. 4 correspond to experimental data and Gaussian fit, respectively, for the two groups of subjects. The horizontal bars indicate to the 95% confidence intervals of the distribution peaks and FWHMs

For painting I, which is assumed to have the most familiar and culturally neutral content, no significant difference in the average selection rate of AGB vs. RGB weight was observed (Fig. 4(a)). However again, the American group of subjects showed a noticeably wider distribution of the selection rates than Chinese subjects. For painting II, which is assumed to be less familiar to subjects with the Chinese cultural background, the distributions of AGB vs. RGB weight differ in both the width and peak position (Fig. 4(c)). For these two paintings, the distributions of the CCT selection rate show almost no cultural differences within the experimental uncertainty (Figs. 4(b) and (d)). However, the average selected CCT is seen to depend on painting content: the fire topic required considerably lower CCT than the Saint George topic.

For painting III, which is assumed to be unfamiliar to both cultural groups, the selection rate distributions show cultural differences for both AGB vs. RGB weight and CCT (Figs. 4(e) and (f), respectively). As compared to Americans, the Chinese group of subjects selected less saturated colors and lower CCT on average. For both $\sigma$ and CCT, the American group of subjects showed wider distributions.

The parameters of the distributions of color quality selection rate for the two cultural groups are summarized in Table 2.

Table 2. Peak positions and FWHMs of the Gaussian distributions of color quality selection rate.

| Objects, tasks | American subjects | | | Chinese subjects | | |
| --- | --- | --- | --- | --- | --- | --- |
| | CCT | AGB vs. RGB weight | | CCT | AGB vs. RGB weight | |
| | | $\sigma_p$ | FWHM | | $\sigma_p$ | FWHM |
| Fruits, "natural" | 3000 | 0.63±0.04 | 0.35±0.07 | 3000 | 0.64±0.04 | 0.33±0.07 |
| Fruits, preference | 3000 | 0.37±0.04 | 0.37±0.07 | 3000 | 0.36±0.04 | 0.21±0.07 |
| Painting I, preference | 2900±260 | 0.38±0.07 | 0.78±0.14 | 3250±370 | 0.39±0.06 | 0.46±0.08 |
| Painting II, preference | 3920±280 | 0.31±0.07 | 0.85±0.23 | 3900±280 | 0.21±0.08 | 0.47±0.15 |
| Painting III, preference | 5950±460 | 0.00+0.07 | 1.29±1.09 | 4260±430 | 0.35±0.07 | 0.59±0.11 |

## 4. Discussion

The above results obtained by using the continuously tunable color rendition engine reveal both similarities and differences in the selection of color quality of illumination by the groups of subjects with different cultural background.

The first important observation is that when selecting "most natural" color rendition properties of illumination for very familiar objects, such as fruits and vegetables, the two groups of subjects exhibited almost identical distributions of selection rates that peak in the vicinity of the high-fidelity pole of the engine ($\sigma = 0.67$; Fig. 3(a)) and have very similar FWHMs of about 0.34. The reason is probably in that when judging on the "most natural" apperance of the familiar objects, the subjects rely on their memory [11], which is a mental capacity that depends on physiological similarity of human brain rather than on cultural differences. These identical distributions of the selection rates also can be considered is a proof for the validation of the experimental approach based on the use of the color rendition engine. Also, this shows no significant racial differences in the physiology of perciving and recalling the appearance of colored objects.

When the task was changed from finding the "most natural" conditions to those characterized as preferred, the average selection rate exhibited a known mismatch with the highest color fidelity or highest general CRI [11,19,20]. In this case, the selections relied on the individual way of judging rather than on memory and the cultural differences in the selection of the color quality of illumination started emerging.

For very familiar objects, such as fruits and vegetables, and even for an artwork with familiar contents (painting I), the average preference selection rate is shifted to more saturated colors ($\sigma_p \approx 0.37$; see Fig. 1(c) and (d)) but shows no dependence on cultural background (Figs. 3(b) and 4(a)). The distribution of the CCT selection rate (Fig. 4(b)) that was obtained for the artwork illumination also showed no significant cultural differences. These observations are in line with the cross-cultural comparison of color emotion for two-color combinations on the like/dislike scale [4]. However, the striking result of our experiments is that despite the similarity in the average judgements on illumination of familiar objects, different cultural groups exhibited statistically significant difference in the width of the distributions for the color saturating ability of illumination. The American group of subjects showed almost twice as wider distributions over AGB vs. RGB weight than the Chinese group for the familiar objects (FWHM of $0.37 \pm 0.07$ vs. $0.21 \pm 0.07$ for fruits and vegetables and $0.78 \pm 0.14$ vs. $0.46 \pm 0.08$ for painting I, respectively). When switching from a very familiar scene (fruits and vegetables) to a less familiar scene (artwork), a higher diversification of the judgements occured within both cultural groups.

Assuming that the racial factor in perceiving colors is to be excludded, the observed difference in the width of the preference selection rate distributions is to be attributed entirely to the difference in the cultural background of the two groups of subjects. The multidimensionality and complexity of the cultural bacground aggravates finding the exact reasons of this phenomenon. However, the problem can be understood in terms of some simplified models. One of such models is G. Hofstede's approach to national cultural dimensions [21]. In particular, we can trace the correlation of our results with the Hofstede's Individualism/Collectivism dimension (IDV), which can be understood as a measure of the integration of individuals into primary groups. The US population scores an IDV of 91, which shows a preference for a loosely-knit social framework, in which individuals are expected to take care of themselves, and implies highly individual decision making. In contrast, the Chinese score an IDV of 20, which infers that Chinese people act more as a group rather than individuals and their decision making might be influenced by group norms (intended behavior of other group members).

Differentiaiton of the level of similarity of an object among different cultural groups of subjects results in more differences of the distribution of color quality preferences. This is illustrated by the results for painting II, which content is assumed to be more familiar to Americans rather than to Chinese (Fig. 4(c)). In this case, not only a smaller FWHM but also a noticeable shift of the peak of the distribution toward more color saturating conditions was observed for Chinese subjects in respect of American subjects. One can speculate that faded colors of sacral artworks are more acceptable for subjects connected to Christian culture, while Chinese are more tending for the compensation of the faded colors by selecting a more saturating blend [22]. Note that the distribution of the selection rates for CCT is still almost independent of cultural backgroud (Fig.4(d)).

The largest cultural differences in the selection of the color quality of illumination were found for painting III, which was assumed to be very unfamiliar for both groups of subjects. Differently from painting II, a reverse shift of the peak of the distribution toward more color dulling conditions was observed for the Chinese in respect of Americans. Meanwhile the Chinese group of subjects exhibited a need for lower cool-white CCTs (~4300 K) in respect of daylight (~6000 K) that the American group selected on average. The width of the distributions for both AGB vs. RGB weight and CCT was larger for the latter group of subjects, probably due to the reasons discussed above. We can imply that these large cultural differences in the selection of the preferences to the color appearance of the unfamiliar artwork are almost entirely determined by aesthetic judgement, which is strongly influenced by geographic, climatic, landscape, architectural, etc. characteristics of environment responsible for the formation of a particular cultural background.

## 5. Conclusions

Using a color rendition engine, which is based on trading-off between color-saturating and color-dulling lights for a particular CCT, we performed a cross-cultural research on the preferences to color quality of illumination of different objects. The approach used in the resesarch was quantified using the statistical indices of color saturation, color fidelity, and color dulling and validated by the selection of culture-background-independent "most natural" color rendition condtions of illumination.

    The preferences to color quality of illumination were investigated for two representative groups of subjects with different cultural background (Chinese and American) that were performing selections by continuously tuning metameric spectra and CCT. For very familiar objects (fruits and vegetables), as well as for a painting, which was assumed to have a familiar content, no cultural differences were found in the peaks of the distribution of the selected blends. For less familiar paintings, the peaks of the distributions were found to differ depending on the content. An unfamiliar painting also showed very different distributions of the preferred color temperature. In all cases, subjects with American cultural background exhibited noticeably wider distributions while selecting preferred blends than Chinese subjects. Such a cultural difference was hypothesized to be linked with the difference in Hofstede's Individualism/Collectivism cultural dimension and the influence of group norms on decision making.

    Our results prove the need for light sources that have the ability to dynamically tune the color quality of illumination. Such sources can facilitate well-being of different individuals both within a single cultural group and within different groups.

**Acknowledgments**

The work at RPI was supported primarily by the Engineering Research Centers Program (ERC) of the National Science Foundation under NSF Cooperative Agreement No. EEC-0812056 and in part by New York State under NYSTAR contract C090145. The work at VU was funded in part by a grant (No. MIP-098/2012) from the Research Council of Lithuania. P. V. also acknowledges a postdoctoral fellowship funded by the European Union Structural Funds project "Postdoctoral Fellowship Implementation in Lithuania."